\documentclass[11pt]{article}

\usepackage{fullpage,epsfig,latexsym,picinpar,amsbsy,amsmath}

\newcommand{\lastcorrections}%
{{\vskip 0.2in
\begin{sloppypar}
    \baselineskip -0.2in
    \tiny\bf\noindent Last Corrections:\\   %%%%% keep this info
xtof: Mar 13 17:17\\
xtof: Sun Mar 15 19:57:51 PST 1998\\
xtof: Mon Mar 16 12:59:39 PST 1998\\
(we worked in the MFCS version here)\\
xtof: Fri Mar 20 23:34:29 PST 1998	\\
xtof: Sat Mar 21 16:18:09 PST 1998\\
marek: Sun Mar 22 19:40:09 PST 1998\\
marek: Wed Mar 25 15:33:59 PST 1998\\
marek: Thu Mar 26 17:27:19 PST 1998\\
xtof: Sun Mar 29 10:16:49 PST 1998\\
xtof: Mon Mar 30 16:49:00 PST 1998\\
xtof: Wed Apr  1 11:41:04 PST 1998\\
marek: Wed Apr  1 17:54:21 PST 1998\\
marek: Thu Apr  2 11:43:33 PST 1998\\
xtof: Fri Apr  3 11:29:40 PST 1998\\
marek: Mon Apr 27 14:41:13 PDT 1998\\
marek: Fri May 1, 1998	\\
xtof: Mon Apr 12 17:48:23 IDT 1999 (including tcs referees comments)\\
marek: Tue Apr 20 10:45:08 PDT 1999
\end{sloppypar}
}}

\newcommand{\margincomment}[1]%
    {{%
      \marginpar{{\tiny\begin{minipage}{0.5in}
                       \begin{flushleft}
                          {#1}
                       \end{flushleft}
                       \end{minipage}
                }}
    }}

\newcommand{\bfx}{{\bf x}}
\newcommand{\bfy}{{\bf y}}
\newcommand{\bfz}{{\bf z}}
\newcommand{\bfr}{{\bf r}}
\newcommand{\bfs}{{\bf s}}

\newcommand{\calA}{{\cal A}}
\newcommand{\calB}{{\cal B}}
\newcommand{\calI}{{\cal I}}

\newcommand{\barT}{{\bar T}}
\newcommand{\barF}{{\bar F}}

\newcommand{\tinyA}{{\mbox{\tiny\it A}}}
\newcommand{\tinyB}{{\mbox{\tiny\it B}}}
\newcommand{\tinyAB}{{\mbox{\tiny\it AB}}}
\newcommand{\tinyC}{{\mbox{\tiny\it C}}}

\newcommand{\VEC}[1]{{\boldsymbol{#1}}}
\newcommand{\BAR}[1]{{\bar{\VEC{#1}}}}
\newcommand{\REV}[1]{\rev{\VEC{#1}}}

\def\stacksymbols #1#2#3#4{\def\theguybelow{#2}
        \def\verticalposition{\lower#3pt}
        \def\spacingwithinsymbol{\baselineskip0pt\lineskip#4pt}
        \mathrel{\mathpalette\intermediary#1}}
\def\intermediary#1#2{\verticalposition\vbox{\spacingwithinsymbol
        \everycr={}\tabskip0pt
        \halign{$\mathsurround0pt#1\hfil##\hfil$\crcr#2\crcr
                \theguybelow\crcr}}}

\newcommand{\rev}[1]{\mathord{\stacksymbols{{\mbox{$\scriptscriptstyle\leftarrow$}}}{#1}{0}{0.5}}}

\newcommand{\braced}[1]{{ \left\{ #1 \right\} }}

\newcommand{\half}{{\mbox{$\frac{1}{2}$}}}
\newcommand{\suchthat}{{\,:\,}}
\newcommand{\ones}[1] {{\left|#1\right|_1}}
\newcommand{\zeros}[1]{{\left|#1\right|_0}}

\newcommand{\corner}[1]%
		{\,\mbox{\epsfig{file=#1corner.eps,width=1ex}}}
\newcommand{\topleft}[1]{{#1}^{\corner{topleft}}}
\newcommand{\topright}[1]{{#1}^{\corner{topright}}}
\newcommand{\bottomleft}[1]{{#1}^{\corner{bottomleft}}}
\newcommand{\bottomright}[1]{{#1}^{\corner{bottomright}}}

\newcommand{\BSM}{{\mbox{\it BSM}}}
\newcommand{\ASM}{{\mbox{\it ASM}}}
\newcommand{\edgeverifier}{{\mbox{\it EV}}}
\newcommand{\perfectmirror}{{\mbox{\it PM}}}

\newcommand{\totalsum}{{\mbox{\large$\Sigma$}}}

\newcommand{\tinymin}{{\mbox{\tiny\it min}}}
\newcommand{\tinymax}{{\mbox{\tiny\it max}}}

\newtheorem{theorem}{Theorem}	\newtheorem{corollary}{Corollary}
\newtheorem{lemma}{Lemma}	\newtheorem{remark}{Remark}

\newcommand{\Qed}{\hspace*{1ex}\hfill\rule{1ex}{1ex}}
\newenvironment{Proof}{\paragraph*{Proof}}{\Qed\medskip}
			
\title{Reconstructing Polyatomic Structures from Discrete X-Rays:\\
		NP-Completeness Proof for Three Atoms}

\begin{document}

%%%%%%%%%%%%%%%%%%%%%%%%%%%%%%%%%%%%%%%%%%%%%%%%%%%%%%%%%%%%%%
%%%%%%%%%%%%%%%%%%%%%%%%%%%%%%%%%%%%%%%%%%%%%%%%%%%%%%%%%%%%%%
%%%%%%%%%%%%%%%%%%%%%%%%%%%%%%%%%%%%%%%%%%%%%%%%%%%%%%%%%%%%%%

\author{Marek Chrobak\thanks{%
	Department of Computer Science,	
        University of California,	
       	Riverside, CA 92521-0304.	
	{Email {\tt marek@cs.ucr.edu}}.
		Research supported by NSF grant CCR-9503498.
                This research was partially conducted when the author
		was visiting International Computer Science Institute
		in Berkeley.
                }			
\and	
	Christoph D\"urr\thanks{%					
                International Computer Science Institute,	
        	1947 Center Street, Suite 600,
	        Berkeley, CA 94704-1198.
	{Email {\tt cduerr@icsi.berkeley.edu}}.
	{URL {\tt http://www.icsi.berkeley.edu/\~{}cduerr/Xray}}.
}}

\date{}

%%%%%%%%%%%%%%%%%%%%%%%%%%%%%%%%%%%%%%%%%%%%%%%%%%%%%%%%%%%%%%
%%%%%%%%%%%%%%%%%%%%%%%%%%%%%%%%%%%%%%%%%%%%%%%%%%%%%%%%%%%%%%
%%%%%%%%%%%%%%%%%%%%%%%%%%%%%%%%%%%%%%%%%%%%%%%%%%%%%%%%%%%%%%

\begin{titlepage}

\maketitle
\pagestyle{plain}				

\begin{abstract}
We address a discrete tomography problem that arises in the study of
the atomic structure of crystal lattices.  A polyatomic structure $T$
can be defined as an integer lattice in dimension $D\ge 2$, whose
points may be occupied by $c$ distinct types of atoms. To ``analyze'' $T$,
we conduct $\ell$ measurements that we call {\em discrete X-rays}.
A discrete X-ray in direction $\xi$ determines the number of
atoms of each type on each line parallel to $\xi$. Given 
$\ell$ such non-parallel X-rays, we wish to reconstruct $T$.

The complexity of the problem for $c=1$ (one atom type) has been
completely determined by Gardner, Gritzmann and Prangenberg
\cite{GaGrPr97a}, who proved that the problem is NP-complete for any
dimension $D\ge 2$ and $\ell\ge 3$ non-parallel X-rays, and that it
can be solved in polynomial time otherwise~\cite{Ryser63}.

The NP-completeness result above clearly extends to any $c\ge 2$, and
therefore when studying the polyatomic case we can assume that $\ell =
2$. As shown in another article by the same authors, \cite{GaGrPr97b},
this problem is also NP-complete for $c\ge 6$ atoms, even for
dimension $D=2$ and axis-parallel X-rays.  The authors of
\cite{GaGrPr97b} conjecture that the problem remains NP-complete for
$c =3,4,5$, although, as they point out, the proof idea in
\cite{GaGrPr97b} does not seem to extend to $c\le 5$.

We resolve the conjecture from \cite{GaGrPr97b} by proving that the
problem is indeed NP-complete for $c\ge 3$ in 2D, even for
axis-parallel X-rays. Our construction relies heavily on some
structure results for the realizations of 0-1 matrices with given row
and column sums.
\end{abstract}

%\lastcorrections


\end{titlepage}

\setcounter{page}{2}

%%%%%%%%%%%%%%%%%%%%%%%%%%%%%%%%%%%%%%%%%%%%%%%%%%%%%%%%%%%%%%
%%%%%%%%%%%%%%%%%%%%%%%%%%%%%%%%%%%%%%%%%%%%%%%%%%%%%%%%%%%%%%
%%%%%%%%%%%%%%%%%%%%%%%%%%%%%%%%%%%%%%%%%%%%%%%%%%%%%%%%%%%%%%

\section{Introduction}
%        ============

The fundamental principle of the {\em transmission electron
microscope\/} (TEM) is very similar to the more familiar optical
microscope: it ``shines'' a focused beam of electrons towards a
specimen, and the transmitted beam is projected onto a phosphor screen,
thereby generating an image. The intensity represents the density and
thickness of the specimen: denser or thicker areas of the specimen
transmit fewer electrons and produce darker areas in the image.  The
development of the TEM in 1930's was necessitated by the limitations
of the optical microscopes, whose magnification and resolution were
insufficient to study the internal structure of organic cells or to
find defects in bulk materials. Recently, new advancements in {\em
high-resolution TEM\/} (HRTEM) led to the development of instruments and
techniques for studying biological molecules and for investigating the
atomic structure of crystals.  In particular, a technique called
QUANTITEM \cite{KSBSKO95,SKSBKO93} allows one to determine the number
of atoms in the atom columns of a crystal in certain directions. Given
these numbers, we wish to reconstruct the structure of the
crystal. This is an example of an algorithmic problem belonging to
{\em discrete tomography}, the area of mathematics and computer
science that deals with inverse problems of reconstructing discrete
density functions from a finite set of projections. The size of
crystals that occur in materials science applications is about
$10^6$ atoms, and, for data sets that large, efficient
reconstruction algorithms would be of great interest.

The problem we address in this paper can be formulated as follows:
Define a {\em polyatomic structure $T$\/} as an integer lattice in
dimension $D\ge 2$, whose cells may be occupied by $c$ distinct types of
atoms.  Each of these cells can be occupied by one atom, or it could be empty.
To ``analyze'' $T$, we conduct $\ell$ measurements that we refer to as
{\em discrete X-rays}. (QUANTITEM uses electron beams, but, following
\cite{GaGrPr97a}, we use a more familiar term ``X-ray'' instead.) A
discrete X-ray in direction $\xi$ determines the number of
atoms of each type on each line parallel to $\xi$. Given such
$\ell$ non-parallel X-rays, we wish to reconstruct $T$.

The complexity of the problem for $c=1$ (one atom type) has been
completely determined by Gardner, Gritzmann and Prangenberg
\cite{GaGrPr97a}, who proved that the problem is NP-hard for any
dimension $D\ge 2$ and $\ell\ge 3$ non-parallel X-rays, and that it
can be solved in polynomial time otherwise~\cite{Ryser63}.

The NP-hardness result above clearly extends to any $c\ge 2$, and
therefore when studying the polyatomic case we can assume that $\ell =
2$. As shown in another article by the same authors, \cite{GaGrPr97b},
this problem is also NP-hard for $c\ge 6$ atoms, even for dimension
$D=2$ and for the axis-parallel X-rays.  The authors of
\cite{GaGrPr97b} conjectured that the problem remains NP-hard for $c
=3,4,5$, and they pointed out that for these values of $c$
``a substantially new
technique will be needed, at least for the case $c=3$''.

We resolve the conjecture from \cite{GaGrPr97b} by proving that the
problem is indeed NP-hard for $c = 3$ (and thus for any larger $c$ as well)
in 2D, even for the orthogonal case, that is, with axis-parallel X-rays.

In the orthogonal case, the problem is equivalent to that of
reconstructing $(c+1)$-valued matrices ($c$ atom types and
``holes'') from the row and column sums for each atom. Without loss
of generality, we can concentrate on square, say $L\times L$, matrices.
Let $\Delta$ be the set of $c$ atom types. For any atom
type $a\in\Delta$, denote by $r_i^a$ (resp. $s_j^a$) the {\em
row-sum\/} (resp. {\em column-sum}) of atom $a$, that is, the number
of atoms of type $a$ in row $i$ (resp. in column $j$). The
vectors $\bfr^a = (r^a_1,\dots,r^a_L)$ and $\bfs^a =
(s^a_1,\dots,s^a_L)$ are referred to, respectively, as the {\em
row-sum vector\/} and the {\em column-sum vector\/} for atom $a$.

A \emph{realization} of the sums $\calI=(\bfr^a,\bfs^a)_{a\in\Delta}$ is
an $L\times L$ matrix $T$ with values from $\Delta\cup \braced{\Box}$,
such that for each atom type $a \in \Delta$
\begin{eqnarray*}
                |\braced{ j \suchthat  T[i,j] = a }| &=& r^a_i
				\quad\quad \forall i = 1,\dots, L
				\\
                |\braced{ i \suchthat  T[i,j] = a }| &=& s^a_j
				\quad\quad \forall j = 1,\dots, L.
\end{eqnarray*}
We say that $\calI$ is \emph{consistent} if it has a realization.

More specifically, we concentrate on the following decision problem:
\begin{quote}
\begin{description}
\samepage
\item[$c$-Color~Consistency~Problem ($c$-CCP)]~
        \begin{description}
        \item[Instance:] row and column sums
                $\calI=(\bfr^a,\bfs^a)_{a\in \Delta}$,
		where $|\Delta| = c$;
        \item[Query:]
                Is $\calI$ consistent?
        \end{description}
\end{description}
\end{quote}

Gardner, Gritzmann and Prangenberg proved in \cite{GaGrPr97b}
that 6-CCP is NP-complete. In this paper we prove that
3-CCP is NP-complete.

If we restrict ourselves further to just one atom (that is, 1-CCP),
the problem becomes
equivalent to the reconstruction of 0-1 matrices from the row and
column sums -- a problem predating the discrete tomography research.
The first efficient reconstruction algorithm was proposed in 1963 by
Ryser \cite{Ryser63}, and a similar algorithm was rediscovered
in 1971 by Chang \cite{Chang71}.  In addition to reconstruction, Ryser
and others studied various structural properties of 0-1 matrices with
given row and column sums, and our construction relies heavily on some
results in this area. Interested readers are referred to an excellent
survey by Brualdi \cite{Brualdi80}.

The general idea of the proof is explained in Section~\ref{sec:
General Idea}. In Section~\ref{sec: 0-1 Matrices}, we review the
structural properties of 0-1 matrices with given row and
column sums that are needed for our proof.
Using these properties, we prove the Skew-Mirror Lemma in
Section~\ref{sec: Skew-Mirror Lemma}.  In Section~\ref{sec: Gadgets},
we construct a number of gadgets, including ``skew mirrors'' and
``edge verifiers'', and we prove that they satisfy the desired
properties.  Finally, in Section~\ref{sec: NP-Completeness Proof}, we
present the complete construction and give the formal NP-completeness
proof.

In addition to the QUANTITEM method, the problem of reconstructing
lattice sets from their projections arises naturally in a variety of
other areas, including statistics, data security, and image
processing. It can also be expressed as a multicommodity flow
problem. We discuss these issues in Section~\ref{sec: Final Comments},
where we also comment on the last unresolved case, $c = 2$.

%%%%%%%%%%%%%%%%%%%%%%%%%%%%%%%%%%%%%%%%%%%%%%%%%%%%%%%%%%%%%%
%%%%%%%%%%%%%%%%%%%%%%%%%%%%%%%%%%%%%%%%%%%%%%%%%%%%%%%%%%%%%%
%%%%%%%%%%%%%%%%%%%%%%%%%%%%%%%%%%%%%%%%%%%%%%%%%%%%%%%%%%%%%%

\section{The General Idea of the Proof} \label{sec: General Idea}
%        =============================

In the proof, we use a reduction from the Vertex Cover problem:
\begin{quote}
\begin{description}
\samepage
\item[Vertex Cover Problem]~
        \begin{description}
        \item[Instance:]
                An undirected graph $G(V,E)$, an
		integer $K$;
         \item[Query:]
                Is there a vertex cover of $G$ of size $K$?
        \end{description}
\end{description}
\end{quote}

Recall that a {\em vertex cover\/} of a graph $G = (V,E)$ is a set $U
\subseteq V$ such that for all $(u,v)\in E$, either $u\in U$ or $v\in
U$. The Vertex Cover problem is well known to be NP-complete (see, for
example, \cite{GarJoh79}).
Let $n = |V|$ and $m = |E|$. We assume, without loss of
generality, that $m,n \ge 1$.

Suppose first that, using some set $\Delta'$ of $d$ atom types, we can
force a unique realization of the form shown in
Figure~\ref{fig: d-atom frame}. We call this realization a {\em frame}.
In the frame, the empty entries form two diagonals, the
{\em main diagonal\/} of length $(m+1)n$, and the {\em side diagonal\/}
of length $mn$. We divide both diagonals into intervals 
of length $n$ that we refer to as {\em mirrors}.
Thus we have two rows of mirrors: $m+1$
mirrors in the main-diagonal, and $m$ mirrors in the side-diagonal.
All other entries are filled with atoms from $\Delta'$.

\medskip

\begin{figwindow}[0,r,
	\epsfig{file = 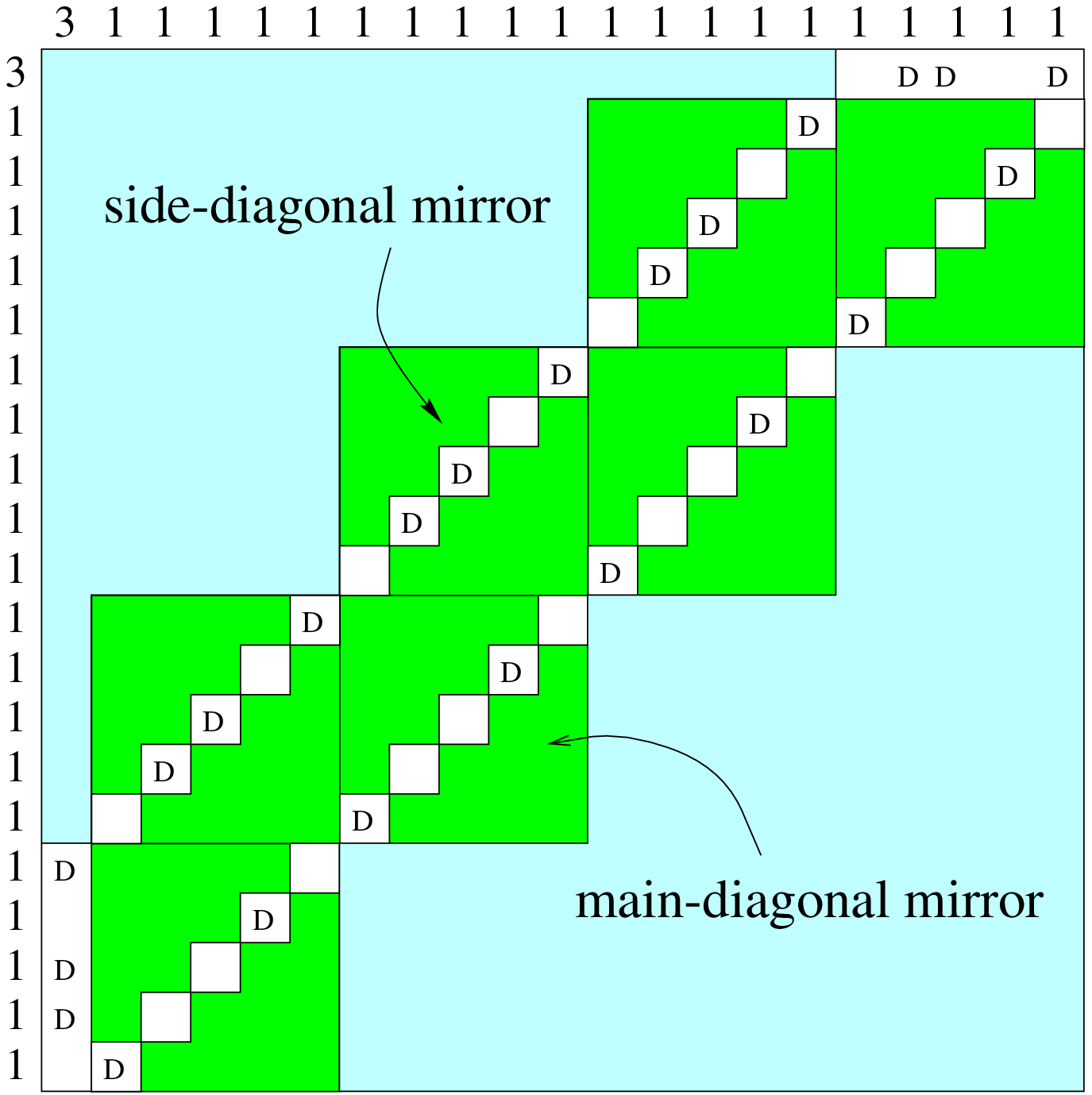,height=2.75in},
\em 	The frame and mirrors for $m=3$.\label{fig: d-atom frame}]
We now add two more atom types $C,D\notin\Delta'$.
Use atom $D$ to create $m$ copies of a candidate vertex cover
$U$ in the following way: The first row and column $D$-sum is $K$ and
all other $D$-sums are 1.  (Figure~\ref{fig: d-atom frame} shows the
$D$-sums.)  Then the pattern of $D$s in each main-diagonal mirror is
the same, and is also the same as the pattern of holes in the
side-diagonal mirrors.  We associate $U$ with this pattern: a vertex $u$
is in $U$ iff the $u$th cell in any side-diagonal mirror is a hole.  We
think of $U$ as a ``beam'' projected onto the last $n$ cells in the
first column, repeatedly reflected in a double-row of mirrors, and
exiting through the last $n$ cells in the first row.

\indent
Finally, we can use atom $C$ to verify that $U$ is indeed a vertex
cover. In order to do so, we convert the $j$th side-diagonal mirror into
an {\em edge verifier\/} for edge $e_j = (u,v)$ (it may be necessary
to add some more rows and columns to the matrix shown in
Figure~\ref{fig: d-atom frame}). Using appropriate sums for atom $C$,
the realization of atoms in $\Delta'$ can be extended to a
realization of all atoms, including $C$, iff either the $u$th cell or
the $v$th cell in side-diagonal mirrors is a hole (and thus, either $u\in
U$ or $v\in U$).
\end{figwindow}

An idea similar to the one described above was used by Gardner,
Gritzmann and Prangenberg \cite{GaGrPr97b} (they used a reduction from
a different problem, not Vertex Cover).  Using 4 atoms they
constructed, in essence, what we call a frame, obtaining the
NP-completeness proof for 6 atoms. In our first attempt to
improve their construction we were able to construct the
frame gadget with only 3 atoms, reducing the total number of
atoms to 5. However, this idea does not work when fewer than 5
atoms are available. As pointed out by \cite{GaGrPr97b}, a new
approach is needed.

The main idea behind our proof is this: Define a partial order
``$\preceq$'' on all $K$-element vertex sets (candidate vertex
covers). The important property of ``$\preceq$'' is that its depth is
polynomial, namely at most $J = K(n-K)+1$
(each strictly increasing chain has
length at most $J$).  Further, ``$\preceq$'' has a unique minimum
element $U^\tinymin$, and a unique maximum element $U^\tinymax$.
Instead of using ``perfect'' mirrors, we use ``skew'' mirrors. These
mirrors have the property that the reflected set is never smaller
(with respect to partial order ``$\preceq$'') than the set
projected onto a skew mirror. These skew mirrors are also ``wobbly''
--- we know that they can reflect the same or a bigger set, but we
cannot control what exactly the reflected set will be.

Now, instead of using $m$ mirrors, we use $mJ$ skew mirrors in the
side-diagonal. They are divided into $J$ segments of $m$ mirrors each.
In each segment, the $j$th skew
mirror is converted into an edge verifier for edge
$e_j$. We ``shine'' $U^\tinymin$ onto the first mirror in the
bottom-left corner, and we make sure that the final set resulting from
all reflections in the top-right corner is $U^\tinymax$.  Since
``$\preceq$'' has depth $J$, there has to be a segment in which all
mirrors reflect the same set $U$. Then the edge verifiers in this
segment will verify that $U$ is indeed a vertex cover.

Why does it help? It turns out that our skew mirrors can be
constructed using only two atom types. Furthermore, the same
atom types can be used to encode the information about
the candidate vertex cover $U$. We use one more
atom type to construct edge verifiers, and thus we only
need three atom types for the whole construction.

%%%%%%%%%%%%%%%%%%%%%%%%%%%%%%%%%%%%%%%%%%%%%%%%%%%%%%%%%%%%%%
%%%%%%%%%%%%%%%%%%%%%%%%%%%%%%%%%%%%%%%%%%%%%%%%%%%%%%%%%%%%%%
%%%%%%%%%%%%%%%%%%%%%%%%%%%%%%%%%%%%%%%%%%%%%%%%%%%%%%%%%%%%%%

\section{0-1 Matrices with Given Row and Column Sums}
\label{sec: 0-1 Matrices}
%        ===========================================

In this section we review some basic results from the literature
on 0-1 matrices with given row and column sums. 

By $\bfx,\bfy,\bfz$ we denote nonnegative integer vectors of length
$p$, for example $\bfx = (x_1,\dots,x_p)$.  The
reconstruction problem for 0-1 matrices with given row and column
sums is equivalent to 1-CCP, and can be stated as follows:
Given $\bfx$ and $\bfy$, is there a 0-1 matrix $T$ that has
$x_i$ 1's in row $i$ and $y_j$ 1's in column $j$, for all $1\le i,j
\le p$?  Again, in this case, a matrix $T$ satisfying these
conditions is called a \emph{realization}, and 
$\bfx,\bfy$ are called \emph{consistent} if they have a realization.

\paragraph*{The structure function.}
Given a $p\times p$ matrix $T$, and integers $0 \leq k,l\leq p$ we
partition $T$ into four submatrices (which may have zero width or
height): $\topleft{T}_{kl}$, $\topright T_{kl}$, $\bottomleft T_{kl}$
and $\bottomright T_{kl}$, defined by the intersections of the first
$k$ rows or the last $p-k$ rows with the first $l$ columns or the last
$p-l$ columns), that is
\[
	T = \left(\begin{array}{cc}
		\topleft{T}_{kl}& 	\topright T_{kl} \\[.8em]
		\bottomleft T_{kl}&	\bottomright T_{kl}
		  \end{array}
	\right).
\]
By $\ones{T}$ and $\zeros{T}$ we
denote the numbers of 1's and 0's in matrix $T$.

For a given instance $\bfx,\bfy$, the {\em structure function\/} $\tau_{kl}$
is defined by
\begin{eqnarray*}
	\tau_{kl} &=& 
	(p-k)(p-l) + \sum_{j=1}^l y_j - \sum_{i=k+1}^p x_i.
\end{eqnarray*}
Then for any arbitrary realization $T$ we have
\begin{align}
&
   \begin{array}{llcc}
	   \tau_{kl} 
   &= 
	   (p-k)(p-l) 
   &+
	   \displaystyle\sum_{j=1}^l y_j 
   &-
	   \displaystyle\sum_{i=k+1}^p x_i
     \\[1em]
   &=
	   \zeros{\bottomright T_{kl}}+\ones{\bottomright T_{kl}}
   &+
	   \ones{\topleft{T}_{kl}}+\ones{\bottomleft T_{kl}}	
   &-
	   \ones{\bottomleft T_{kl}}-\ones{\bottomright T_{kl}}
   \end{array}
\nonumber\\
&
  \begin{array}{ll}
     	   \phantom{ \tau_{kl} }
  &=	   \zeros{\bottomright T_{kl}} + \ones{\topleft{T}_{kl}}.
  \end{array}
  \label{eqn: tau equation}
\end{align}

\paragraph*{Consistent sums.}
We now show that, using the structure function, it is possible to characterize
consistent sums.  An integer vector $\bfz=(z_1,\ldots,z_p)$ will be
called \emph{monotone} if $z_1\leq \ldots \leq z_p$.

%%%%%%%%%%%%%%%%%%%%%%%%%%%%%%%%%%%%%%%%%%%

\begin{lemma} {\rm \cite{Brualdi80}\/}		\label{lem: consistency}
  Monotone vectors $\bfx,\bfy$ are consistent if and only if
  $\tau_{kl}\ge 0$ for all $k,l = 1,\dots,p$.
\end{lemma}

The implication $(\Rightarrow)$ in Lemma~\ref{lem: consistency}
follows directly from Equation~(\ref{eqn: tau equation}).  The
implication $(\Leftarrow)$ can be proven constructively by giving an
algorithm that produces a realization $T$ for any pair $(\bfx,\bfy)$ for
which the structure function is non-negative. (See \cite{Brualdi80}
for details.) It is also not hard to see that
Lemma~\ref{lem: consistency} can be derived from the
Max-Flow-Min-Cut theorem for network flows. 

%%%%%%%%%%%%%%%%%%%%%%%%%%%%%%%%%%%%%%%%%%%

\paragraph*{Decomposed realizations.}
We say that $T$ is {\em $(k,l)$-decomposed\/} if $\topleft{T}_{kl}$
consists only of 0's and $\bottomright{T}_{kl}$ consists only of 1's.
The lemma below follows immediately from Equation~(\ref{eqn: tau
equation}), and it will play a major role in this paper.  Note that in
this lemma, as in the definition of $\tau$, we do not require the
projections $\bfx,\bfy$ to be monotone.

%%%%%%%%%%%%%%%%%%%%%%%%%%%%%%%%%%%%%%%%%%%

\begin{lemma} {\rm \cite{Brualdi80}\/}	 	\label{lem: decomposability}
Suppose that $T$ is a realization of $\bfx,\bfy$, and let $0\le k,l\le p$.
Then $\tau_{kl}=0$ if and only if $T$ is $(k,l)$-decomposed.
\end{lemma}
\begin{remark}				\label{rmk: decomposed as well}
   Lemma~{\rm \ref{lem: decomposability}} implies that if just one
   realization of $\bfx,\bfy$ is $(k,l)$-decomposed, then {\em all\/}
   realizations are $(k,l)$-decomposed as well.
\end{remark}

%%%%%%%%%%%%%%%%%%%%%%%%%%%%%%%%%%%%%%%%%%%%%%%%%%%%%%%%%%%%%%
%%%%%%%%%%%%%%%%%%%%%%%%%%%%%%%%%%%%%%%%%%%%%%%%%%%%%%%%%%%%%%
%%%%%%%%%%%%%%%%%%%%%%%%%%%%%%%%%%%%%%%%%%%%%%%%%%%%%%%%%%%%%%

\section{The Skew-Mirror Lemma}\label{sec: Skew-Mirror Lemma}
%        ===========================================

%%%%%%%%%%%%%%%%%%%%%%%%%%%%%%%%%%%%%%%%%%%%%%%%%%%%%%%%%%%%%

\paragraph*{0-1 Vectors and minorization.}
We use Greek letters $\VEC\alpha,\VEC\beta,\ldots$ for 0-1 vectors of
length $p$, say $\VEC\alpha = (\alpha_1,\dots,\alpha_p)$.  The
\emph{complement} $\BAR\alpha$ of $\VEC\alpha$ is $\bar\alpha_i = 1 -
\alpha_i$ for all $i= 1,\dots,p$, and the \emph{reverse} $\REV\alpha$
is $\rev\alpha_i = \alpha_{p-i+1}$ for $i = 1,\dots,p$.

We say that $\VEC\alpha$ \emph{minorizes} $\VEC\beta$, denoted
$\VEC\alpha\preceq\VEC\beta$, if
\begin{eqnarray*}
      \sum_{i=1}^k \alpha_i &\le&  \sum_{i=1}^k \beta_i
        \quad\quad \forall\; k = 1,\dots, p.
\end{eqnarray*}
By straightforward verification, ``$\preceq$'' is a partial order.
We also write $\VEC\alpha\prec\VEC\beta$
if $\VEC\alpha\preceq\VEC\beta$ and $\VEC\alpha\neq\VEC\beta$.

The \emph{total sum} of a 0-1 vector $\VEC\alpha$ is
$\totalsum\VEC\alpha=\sum_{i=1}^p \alpha_i$. If $\VEC\alpha$,
$\VEC\beta$ are two 0-1 vectors with
equal total sums, then the definitions above imply directly the
following equivalences:
\begin{eqnarray*}
\VEC\alpha \preceq \VEC\beta
        \quad\iff\quad
        \BAR\alpha \succeq \BAR\beta
        \quad\iff\quad
        \REV\alpha \succeq \REV\beta.
\end{eqnarray*}
%

%%%%%%%%%%%%%%%%%%%%%%%%%%%%%%%%%%%%

An important property of the minorization relation is that it is
``shallow'', that is its depth is only polynomial (unlike, for
example, the lexicographic order).  The next lemma gives a more
accurate estimate on the depth of ``$\preceq$''.

%%%%%%%%%%%%%%%%%%%%%%%%%%%%%%%%%%%%

\begin{lemma}\label{lem: majorization is shallow}
Suppose that we have a strictly increasing sequence of 0-1 vectors
\begin{eqnarray*}
\VEC\alpha^1\; \prec \; \VEC\alpha^2 \; \prec \dots \; \prec \;
	\VEC\alpha^q,
\end{eqnarray*}
with total sums $\totalsum\VEC\alpha^i = t$ for each $i$.
Then $q\le t(p-t)+1$.
\end{lemma}

\begin{Proof}
To each 0-1 vector $\VEC\alpha$ assign the number $\|\VEC\alpha\|$
defined by
\begin{eqnarray*}
        \|\VEC\alpha\| &=& \sum_{j=1}^p (p-j+1)\alpha_j \;=\;
		\sum_{k=1}^p\sum_{i=1}^k\alpha_i.
\end{eqnarray*}
If $\VEC\alpha\prec\VEC\beta$ then $\sum_{i=1}^k\alpha_i\le
\sum_{i=1}^k\beta_i$ for $k=1,\dots,p$, and this inequality must be
strict for at least one $k$.  We conclude that
$\VEC\alpha\prec\VEC\beta$ implies $\|\VEC\alpha\| < \|\VEC\beta\|$.

Now, by the argument above, the numbers $\|\VEC\alpha^i\|$ are strictly
increasing. Therefore
\begin{eqnarray*}
q &\le& \|\VEC\alpha^q\| - \|\VEC\alpha^1\| + 1 \\
  &\le& \sum_{k=p-t+1}^p k - \sum_{k=1}^t k + 1 \\
  &=& t(p-t)+1,
\end{eqnarray*}
completing the proof.
\end{Proof}

%%%%%%%%%%%%%%%%%%%%%%%%%%%%%%%%%%%%%%%%%%%%%%%%%%%%%%%%%%%%%

\paragraph*{The 0-1 skew mirror.}
The lemma below deals with a special instance of the reconstruction
problem for 0-1 matrices, in which the row sum vector $\bfx$ is
determined by a 0-1 vector $\VEC\alpha$, and the column-sum vector
$\bfy$ is determined by a 0-1 vector $\VEC\beta$.

Given a 0-1 vector $\VEC\sigma$ of length $p$, we associate with
$\VEC\sigma$ a $p\times p$ {\em perfect mirror\/} matrix
$\perfectmirror_{\VEC\sigma}$
defined by
\begin{eqnarray*}
\perfectmirror_{\VEC\sigma}[i,j] &=&
	\left\{\begin{array}{lcl}
		0 & \quad & \mbox{\rm for\ } i+j \le p \\
		\sigma_i & \quad & \mbox{\rm for\ } i+j = p+1 \\
		1 & \quad & \mbox{\rm for\ } i+j \ge p+2. \\
	\end{array}\right.
\end{eqnarray*}
In a perfect-mirror matrix the cells on the main diagonal $i+j = p+1$,
counted from top down, contain $\VEC\sigma$, while all cells above it
are 0, and all cells below it are 1 (see Figure~\ref{fig:
mirror}b). From Lemma~\ref{lem: decomposability} we immediately obtain
the following corollary.

%%%%%%%%%%%%%%%%%%%%%%%%%%%%%

\begin{corollary}\label{cor: perfect mirror}
Let $\VEC\sigma$ be a 0-1 vector of length $p$.
Then $\perfectmirror_{\VEC\sigma}$ is a realization of
vectors $\bfx,\bfy$ if and only if for each $k = 1,\dots,p$, 
{\rm \/(a)\/}
$\tau_{k,p-k} = 0$, and
{\rm \/(b)\/}
$\tau_{k,p+1-k} = 0$ iff $\sigma_k = 0$.
\end{corollary}

Note that Corollary~\ref{cor: perfect mirror} together with
Remark~\ref{rmk: decomposed as well} implies that if
$\perfectmirror_{\VEC\sigma}$ is a realization of $\bfx,\bfy$ then it
is unique.

%%%%%%%%%%%%%%%%%%%%%%%%%%%%%

\begin{lemma}  \label{lem: skew mirror}
Let $\VEC\alpha,\VEC\beta$ be two 0-1 vectors of length $p$, and let
$\bfx,\bfy$ be row and column sums defined by $x_i = i -\alpha_i$ and
$y_i = i-\beta_i$, for $i = 1,\dots,p$. Then
\begin{description}
\item{(a)} Vectors $\bfx,\bfy$ are consistent iff
	$\totalsum\REV\alpha= \totalsum\VEC\beta$ and
	$\REV\alpha\succeq\VEC\beta$.
\item{(b)} Suppose that $\bfx,\bfy$ are consistent. Then
	$\REV\alpha = \VEC\beta$ iff the unique
	realization of $\bfx,\bfy$ is $\perfectmirror_{\bar{\VEC\alpha}}$.
\end{description}
\end{lemma}

%%%%%%%%%%%%%%%%%%%%%%

\begin{figure}[tb]
\centerline{
	\begin{tabular}{c@{\hspace{2cm}}c}
  \epsfig{file=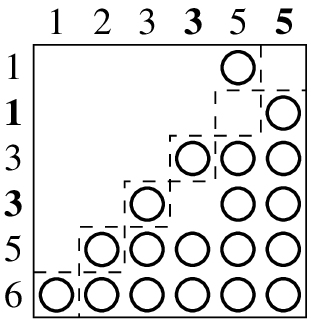,width=3cm}
  &
  \epsfig{file=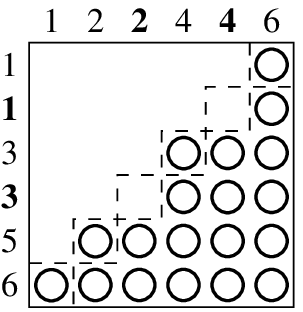,width=3cm}
	\\
	\hspace*{2em}(a)	&
	\hspace*{2em}(b)
	\end{tabular}
}
\caption{\em\small  
(a) A realization of $\bfx$, $\bfy$ for $\VEC\alpha = 010100$ and
$\VEC\beta = 000101$. Disks represent $1$'s. The row and column
sums that correspond to $0$'s in $\VEC\alpha$ and $\VEC\beta$ are
shown in bold.
(b) Perfect mirror $\perfectmirror_{\VEC\sigma}$ for $\VEC\sigma = 101011$.}
\label{fig: mirror}
\end{figure}

%%%%%%%%%%%%%%%%%%%%%%

Lemma~\ref{lem: skew mirror} is illustrated in Figure~\ref{fig: mirror}.
In Figure~\ref{fig: mirror}(a),
we have $\VEC\alpha = 010100$, and $\VEC\beta =  000101$
(we write 0-1 vectors as binary strings, for simplicity),
for which the corresponding row sums and columns sums
are $\bfx = (1,1,3,3,5,6)$ and $\bfy = (1,2,3,3,5,5)$. 
Note that $\totalsum\REV\alpha = \totalsum\VEC\beta = 3$, and
that $\REV\alpha\succeq\VEC\beta$, implying that
$\bfx$ and $\bfy$ are consistent. One realization of
$\bfx$, $\bfy$ is shown in Figure~\ref{fig: mirror}(a).
In Figure~\ref{fig: mirror}(b), $\VEC\alpha$ and $\bfx$ are the same as in (a).
Since $\bfy = (1,2,2,4,4,6)$ corresponds to
$\VEC\beta = 001010 = \REV\alpha$, vectors $\bfx$, $\bfy$
have a unique realization that is a perfect
mirror $\perfectmirror_{\VEC\sigma}$,
for $\VEC\sigma = 101011 = \bar{\VEC\alpha}$.

%%%%%%%%%%%%%%%%%%%%%%%%%%

\begin{Proof}
Vectors $\bfx,\bfy$ are monotone, so we can use Lemma~\ref{lem: consistency}.
We start by computing the structure function for $\bfx$ and $\bfy$:
\begin{eqnarray}
\tau_{kl}
&=& 	(p-k)(p-l) 
	+ \sum_{j=1}^l y_j
	- \sum_{i=k+1}^p x_i 
						\nonumber \\
&=&	(p-k)(p-l) 
	+ \sum_{j=1}^l	 ( j-\beta_j)
	- \sum_{i=k+1}^p ( i-\alpha_i)
						\nonumber \\
&=& 	  \sum_{i=k+1}^p \alpha_i 
	- \sum_{j=1}^l	 \beta_j 
	+ \half[\,2(p-k)(p-l)-p(p+1)+k(k+1)+l(l+1)\,]
							\nonumber \\
&=& 	  \sum_{i=1}^{p-k} \rev\alpha_i 
	- \sum_{j=1}^l	 \beta_j 
	+ \half(p-k-l-1)(p-k-l).
	\label{eqn: tau for alpha beta}
\end{eqnarray}
\noindent
Now we are ready to prove Part~(a). We prove the two implications
separately.

\begin{description}
\item[$(\Rightarrow)$] For any $l = 0,\dots,p$, using
Equation~(\ref{eqn: tau for alpha beta}) with $k=p-l$, we get that
$\tau_{p-l,l}\ge 0$ implies $\sum_{i=1}^{l} \rev\alpha_i -
\sum_{j=1}^l \beta_j\geq 0$.  Thus $\REV\alpha\succeq\VEC\beta$.
Moreover, $\bfx$ and $\bfy$ have equal total sums, if and only if
$\totalsum\VEC\alpha = \totalsum\VEC\beta$.
\item[$(\Leftarrow)$]
Assume that $\REV\alpha\succeq\VEC\beta$. We consider two cases, when
$k+l\le p$ and $k+l\ge p+1$.

Suppose first that $k+l\leq p$. From
Equation~(\ref{eqn: tau for alpha beta}) we have
\begin{eqnarray*}
\tau_{kl} 
&=& 	\sum_{i=1}^l \rev\alpha_i 
	+ \sum_{i=l+1}^{p-k} \rev\alpha_i 
	- \sum_{j=1}^l	 \beta_j 
	+ \half(p-k-l-1)(p-k-l)\\
&\ge&0,
\end{eqnarray*}
because $\sum_{i=1}^l \rev\alpha_i - \sum_{j=1}^l\beta_j \ge 0$,
and $(p-k-l-1)(p-k-l) \ge 0$.

Suppose now that $k+l\ge p+1$. From
Equation~(\ref{eqn: tau for alpha beta}) we have
\begin{eqnarray*}
\tau_{kl}
&=&     \sum_{i=1}^l \rev\alpha_i
        - \sum_{i=p-k+1}^{l} \rev\alpha_i
        - \sum_{j=1}^l   \beta_j
        + \half(p-k-l-1)(p-k-l)\\
&=&     \sum_{i=1}^l \rev\alpha_i
        - \sum_{j=1}^l   \beta_j
        - \sum_{i=p-k+1}^{l} \rev\alpha_i
        + \half(k+l+1-p)(k+l-p)\\
&=&     \sum_{i=1}^l \rev\alpha_i
        - \sum_{j=1}^l   \beta_j
        + \half\sum_{i=p-k+1}^{l} (k+l+1-p-2\rev\alpha_i) \\
&\ge&   0,
\end{eqnarray*}
because $\sum_{i=1}^l \rev\alpha_i - \sum_{j=1}^l \beta_j \ge 0$,
and $k+l+1-p-2\rev\alpha_i\ge 2-2\rev\alpha_i\ge 0$.
\end{description}

Now we prove Part~(b).  By Corollary~\ref{cor: perfect mirror}
and Equation~(\ref{eqn: tau for alpha beta}), a realization
of $\bfx,\bfy$ is a perfect mirror $\perfectmirror_{\VEC\sigma}$,
for some $\VEC\sigma$, if and only if
$\REV\alpha = \VEC\beta$. Thus it is sufficient to show that
$\REV\alpha = \VEC\beta$ implies that $\VEC\sigma = \bar{\VEC\alpha}$.
This follows by simple verification of row sums.
\end{Proof}

%%%%%%%%%%%%%%%%%%%%%%%%%%%%%%%%%%%%%%%%%%%%%%%%%%%%%%%%%%%%%%
%%%%%%%%%%%%%%%%%%%%%%%%%%%%%%%%%%%%%%%%%%%%%%%%%%%%%%%%%%%%%%
%%%%%%%%%%%%%%%%%%%%%%%%%%%%%%%%%%%%%%%%%%%%%%%%%%%%%%%%%%%%%%

\section{Some Useful Gadgets} 	\label{sec: Gadgets}
%        ===================

\begin{figure}[htb]
\centerline{\epsfig{file=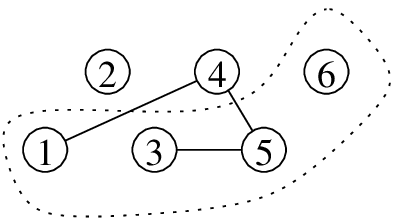}}
\caption{Example of a vertex cover of size $K=4$.}
						\label{fig:graph}
\end{figure}

Recall that $G,K$ is the given instance of Vertex Cover, where $G =
(V,E)$, $n = |V|$ and $m = |E|$. Without loss of generality we can
assume that $0\le K\le n$.  All our examples will refer to the graph
of Figure~\ref{fig:graph}. In this figure, $G$ has $n=6$ vertices,
$m = 3$ edges, and set $\braced{1,3,5,6}$ is a vertex cover
with $K=4$ vertices.
(Clearly, this is not a minimum-size vertex cover. We use this example
only to illustrate certain features of our construction.)

%%%%%%%%%%%%%%%%%%%

%%%%%%%%%%%%%%%%%%%

The purpose of this section is to introduce two $(n+2)\times
(n+2)$-instances of 3-CCP called the
\emph{beige skew mirror} and the \emph{edge verifier}.  They will be
used later in the NP-completeness proof.

Throughout the rest of the paper we will use capital letters $A,B,C$
to denote the three atom types, and we will sometimes refer to these
types as colors: {\it \underline{A}zure}, {\em \underline{B}eige}, and
{\it \underline{C}yan}.

\begin{figure}[htb]
\centerline{\begin{tabular}{c@{\hspace*{1cm}}c}
        \epsfig{file=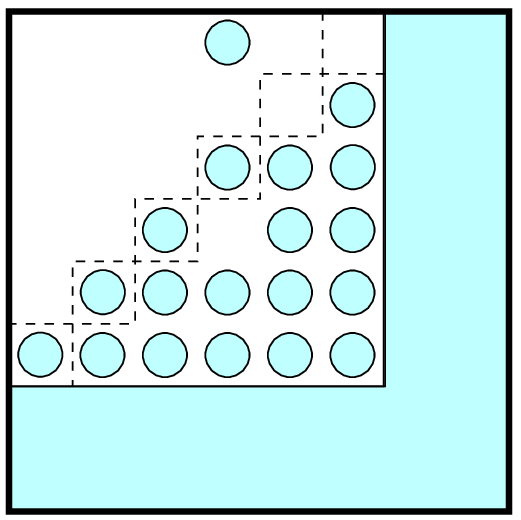,width=3cm}&\epsfig{file=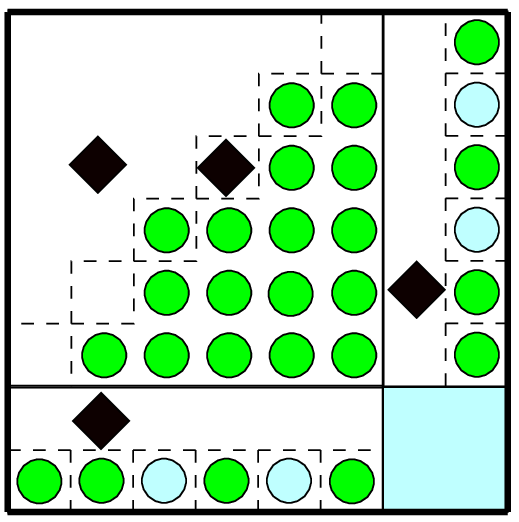,width=3cm}\\
        (a)        &       (b)
            \end{tabular}
        \begin{tabular}{l}
        A  = \epsfig{file=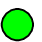,width=.9em},     \\
        B  = \epsfig{file=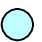,width=.9em} and   \\
        C  = \epsfig{file=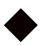,width=.9em}.
        \end{tabular}
}
\caption{\em\small (a) A realization of $\BSM(010100,000011)$.
	(b) A realization of $\edgeverifier(010100,001010,(3,5))$;
	it verifies that the vertex set $\{1,3,5,6\}$ covers the edge $(3,5)$.
	Solid filled regions represent entries that are
	filled with beige atoms independently of the
	parameters of $\BSM$ and $\edgeverifier$. }
\label{fig: pbs}
\end{figure}

\paragraph*{Beige skew mirror.}
Given two 0-1 vectors $\VEC\alpha,\VEC\beta$ of length $n$, we define
the {\em beige skew mirror\/} as an $(n+2)\times (n+2)$ instance of
3-CCP, $\BSM(\VEC\alpha,\VEC\beta) = (\bfx^\tinyB,\bfy^\tinyB)$, with
the following row and column sums:
\begin{eqnarray*}
\begin{array}{lclclcll}
  x^\tinyB_i     &\;=\;& i-\alpha_i + 2
    & & y^\tinyB_i &\;=\;&  i-\beta_i + 2&
                        \quad \mbox{for\ } i = 1,\dots,n\\
  x^\tinyB_i &=& n+2  & &
                y^\tinyB_i &=& n+2 &
			\quad \mbox{for\ } i = n+1,n+2.
\end{array}
\end{eqnarray*}
The azure and cyan sums are zero. Figure~\ref{fig: pbs}(a) shows an
example of a beige skew mirror.

%%%%%%%%%%%%%%%%%%%%%%%%%

\begin{lemma}	\label{lem: beige skew mirror}
Let $\VEC\alpha, \VEC\beta$ be two 0-1 vectors of length $n$.
Then $\BSM(\VEC\alpha,\VEC\beta)$ is consistent if and only if
$\totalsum\VEC\alpha =\totalsum\VEC\beta$ and $\REV\alpha \succeq \VEC\beta$.
\end{lemma}

\begin{Proof}
By definition, any realization of $\BSM(\VEC\alpha,\VEC\beta)$ has its
last 2 rows and last 2 columns completely filled with beige
atoms. Define
\begin{eqnarray*}
	x_i &=& x^\tinyB_i-2
		 \quad=\quad i-\alpha_i,\\
	y_i &=& y^\tinyB_i - 2
		 \quad=\quad i-\beta_i,
\end{eqnarray*}
where $i = 1,\dots,n$. Then $\BSM(\VEC\alpha,\VEC\beta)$ is consistent
if and only if the instance $(\bfx,\bfy)$ of 1-CCP is consistent. Applying
Lemma~\ref{lem: skew mirror}, we obtain that
$\BSM(\VEC\alpha,\VEC\beta)$ is consistent if and only if
$\totalsum\VEC\alpha =\totalsum\VEC\beta$ and
$\REV\alpha \succeq \VEC\beta$.
\end{Proof}

%%%%%%%%%%%%%%%%%%%%%%%%%%%%%%

%%%%%%%%%%%%%%%%%%%%%%%%%%%%%%

\paragraph*{Azure skew mirror.}
Given two 0-1 vectors $\VEC\gamma,\VEC\delta$ of length $n$, we define
the {\em azure mirror\/} as an $(n+2)\times (n+2)$ instance of 3-CCP,
$\ASM(\VEC\gamma,\VEC\delta) = (\bfx^\tinyA,\bfy^\tinyA,\,
\bfx^\tinyB,\bfy^\tinyB)$, with the following row and column sums:
\begin{eqnarray*}
\begin{array}{*2{lcl@{\hspace*{2em}}}l}
  x^\tinyA_i   &=&  i    & 
        y^\tinyA_i   &=& i             &\quad \mbox{for $i =1,\dots,n$}\\
  x^\tinyA_{n+1} &=& 0   & 
        y^\tinyA_{n+1} &=& 0	\\ 
  x^\tinyA_{n+2} &=& K   & 
        y^\tinyA_{n+2} &=& K
\\[2em]
  x^\tinyB_i  &\;=\;& \gamma_i &
        y^\tinyB_i &\;=\;& \delta_i&
                        \quad \mbox{for $i = 1,\dots,n$}\\
  x^\tinyB_{n+1} &=& 2  & 
        y^\tinyB_{n+1} &=& 2
\\
  x^\tinyB_{n+2} &=& n-K+2    & 
        y^\tinyB_{n+2} &=&  n-K+2.
\end{array}
\end{eqnarray*}
%
%>The cyan sums in $\ASM(\VEC\gamma,\VEC\delta)$ are assumed to be zero.
The cyan sums are zero.

%%%%%%%%%%%%%%%%%%%%%%%%%%%%%%

\begin{lemma}   \label{lem: azure skew mirror}
Let $\VEC\gamma,\VEC\delta$ be two 0-1 vectors of length $n$ such that
$\totalsum\VEC\gamma = \totalsum\VEC\delta = n-K$. 
Then $\ASM(\VEC\gamma,\VEC\delta)$ is consistent if and only if
$\VEC\gamma\succeq\REV\delta$.
\end{lemma}

\begin{Proof}
We claim that each realization of $\ASM(\VEC\gamma,\VEC\delta)$ has
beige atoms on positions:
\begin{eqnarray*}
\begin{array}{l@{\hspace*{2em}}l}
        \braced{( i, n+2) : \gamma_i = 1 }
                        &\mbox{(last column)}   \\
        \braced{ ( n+2, i) : \delta_i = 1 }
                        &\mbox{(last row)} \\
        \braced{(n+1,n+1), (n+1,n+2), (n+2,n+1), (n+2,n+2)}.
                        &\mbox{(lower right $2\times 2$ corner)}\\
\end{array}
\end{eqnarray*}
Since $x^\tinyB_{n+2} = n-K+2$, and there are exactly $n-K+2$ non-zero
beige column sums, the beige atoms are determined uniquely, as shown
above.  Similarly, the beige atoms in the last column are determined
uniquely. The last yet unallocated beige atom must be at $(n+1,n+1)$.

We now examine azure atoms.  Row $n+1$ and column $n+1$ have no azure
atoms.  In row $n+2$ azure atoms are forced to be in columns $i$ for
which $\delta_i = 0$, since all other positions are occupied by beige
atoms. Similarly, in column $n+2$ azure atoms are in rows $i$ for
which $\gamma_i = 0$.

Let  $\bfx,\bfy$ be the following row and column sum vectors:
\begin{eqnarray*}
        x_i &=& i-\bar\gamma_i\\
        y_i &=& i-\bar\delta_i,
\end{eqnarray*}
for $i = 1,\dots,n$.  Then $\ASM(\VEC\gamma,\VEC\delta)$ is consistent
iff the instance $(\bfx,\bfy)$ of 1-CCP is consistent.
By Lemma~\ref{lem: skew mirror},
this is equivalent to $\rev{\BAR\gamma}\succeq \BAR\delta$, or
$\VEC\gamma\succeq\REV\delta$, completing the proof.
\end{Proof}

%%%%%%%%%%%%%%%%%%%%%%%%%%%%%%

\paragraph*{Edge verifier.}
For two 0-1 vectors $\VEC\gamma,\VEC\delta$ of length $n$,
and for an edge $e=(u,v)$ (with $u<v$) we define the
{\em edge verifier\/} for $e$, as a $(n+2)\times (n+2)$ instance of 3-CCP,
\begin{eqnarray*}
\edgeverifier(\VEC\gamma,\VEC\delta,e) 
&=& 
	(\bfx^\tinyA,\bfy^\tinyA,\,
	 \bfx^\tinyB,\bfy^\tinyB,\,
	 \bfx^\tinyC,\bfy^\tinyC),
\end{eqnarray*}
where the azure and beige sums are exactly the same as in the azure
skew mirror $\ASM(\VEC\gamma,\VEC\delta)$, and the cyan sums are:
\begin{eqnarray*}
\begin{array}{lcl@{\hspace*{2em}}lcl}
  x^\tinyC_u   &=& 2   & 
        y^\tinyC_{n-u+1}   &=& 1           \\
  x^\tinyC_v   &=& 1   & 
        y^\tinyC_{n-v+1}   &=& 2           \\
  x^\tinyC_{n+1} &=& 1 & 
        y^\tinyC_{n+1} &=& 1.
\end{array}
\end{eqnarray*}

%%%%%%%%%%%%%%%%%%%%%%%%%%%%%

\begin{lemma}	\label{lem: edge verifier}
Let $\VEC\gamma$ be a 0-1 vector of length $n$, and
$e=(u,v)$ (with $u<v$) be an edge of $G$.  Then
$\edgeverifier(\VEC\gamma,\REV\gamma,e)$ is consistent if and only if
$\totalsum\VEC\gamma = n-K$ and either $\gamma_u=0$ or $\gamma_v=0$.
\end{lemma}

Lemma~\ref{lem: edge verifier} has the following interpretation: if we
associate with $\VEC\gamma$ the vertex set $U=\braced{u :
\gamma_u=0}$, then $\edgeverifier(\VEC\gamma,\REV\gamma,e)$ is
consistent if and only if at least one endpoint of edge $e$ belongs to
$U$. See Figure~\ref{fig: pbs}(b) for an example of 
an edge verifier.

\begin{Proof} 
We can assume that $\totalsum\VEC\gamma = n-K$.
By Lemma~\ref{lem: azure skew mirror},
$\ASM(\VEC\gamma,\REV\gamma)$ is consistent. Furthermore, by Part~(b)
of Lemma~\ref{lem: skew mirror}, for any $1\le i,j\le n$, a
realization $T$ of $\ASM(\VEC\gamma,\REV\gamma)$, satisfies: $T[i,j] =
\Box$ for $i+j\le n$, $T[i,j] = A$, for $i+j\ge n+2$, and for
$i+j=n+1$ we have the following equivalence: $T[i,j] = \Box$ iff
$\gamma_i = 0$.

If $\edgeverifier(\VEC\gamma,\REV\gamma,e)$ is consistent, we can
extend $T$ to a realization of $\edgeverifier(\VEC\gamma, \REV\gamma,
e)$, and consider the positions of cyan atoms.  The position
$(n+1,n+1)$ contains a beige atom and $(v,n-u+1)$ an azure atom. This
leaves these possible positions for the cyan atoms:
\[
	\begin{array}{r@{\hspace*{2em}}r@{\hspace*{2em}}r}
	(u,n-v+1)	& (u,n-u+1)	& (u,n+1)	\\
	(v,n-v+1)	& 		& (v,n+1)	\\
	(n+1,n-v+1)	& (n+1,n-u+1).
	\end{array}
\]

We claim that either $T[u,n-u+1] = C$ or $T[v,n-v+1] = C$.  For
otherwise, the cyan row sums $x^\tinyC_u = 2$ and
$x^\tinyC_v = 1$ force $T[u,n+1] = T[v,n+1] = C$, contradicting
$y^\tinyC_{n+1} = 1$.

In summary, we get that $\edgeverifier(\VEC\gamma,\REV\gamma,e)$ is
consistent iff one of $\gamma_u$, $\gamma_v$ equals 0.
\end{Proof}

%%%%%%%%%%%%%%%%%%%%%%%%%%%%%%%%%%%%%%%%%%%%%%%%%%%%%%%%%%%%%%
%%%%%%%%%%%%%%%%%%%%%%%%%%%%%%%%%%%%%%%%%%%%%%%%%%%%%%%%%%%%%%
%%%%%%%%%%%%%%%%%%%%%%%%%%%%%%%%%%%%%%%%%%%%%%%%%%%%%%%%%%%%%%

\section{The Proof of NP-Completeness} \label{sec: NP-Completeness Proof}
%        =============================

In this section we give the overall reduction.  We will define an
instance of 3-CCP, which has a solution if and only if there is a
vertex cover.  Its matrix can be divided into regular quadratic
\emph{blocks} and we will show that in every realization some blocks
will be realizations of particular beige skew mirrors, or of
particular edge verifiers, while the remaining blocks are filled
either with beige or azure atoms.  For this purpose we partition the
matrix into regions of different shapes, which we call
\emph{mirrors, gutters, screens} and \emph{frames}, and we show some
particular properties of them.  Then we conclude that these properties
imply the existence of a vertex cover.

%%%%%%%%%%%%%%%%%%%%%%%%%%%%%%%%%%%%%%%%%%%%%%%%%%%%%%%%%%%%%%

\subsection{The Reduction} \label{sec: The reduction}
%           -------------

Recall that $G = (V,E),K$ is the given instance of Vertex Cover,
where $|V| = n$, $|E| =m$ and $0\le K\le n$.
Define $J = K(n-K)+1$ and $L = (mJ+1)(n+2)$.  We now show how to
map $G,K$ into an $L \times L$ instance of 3-CCP
\begin{eqnarray*}
\calI &=& (\bfr^\tinyA, \bfs^\tinyA,\, \bfr^\tinyB, \bfs^\tinyB,\,
                \bfr^\tinyC, \bfs^\tinyC).
\end{eqnarray*}
To specify the sums in $\calI$, it is convenient to view $L\times L$-matrices
as being partitioned into $(mJ+1)^2$ submatrices of size
$(n+2)\times (n+2)$, called \emph{blocks}.  A row or
column index is then defined by its \emph{block index} $a = 0,\dots,mJ$ and
\emph{offset} $i = 1,\dots,n+2$. For $a \neq mJ$ the azure and beige
sums are:
\begin{eqnarray*}
r^\tinyA_{a(n+2)+i} \;=\; s^\tinyA_{a(n+2)+i} &=& 
	(mJ-a-1)(n+2) +
	\left\{\begin{array}{lcl}
		i &\quad& i = 1,\dots,n \\
		0     & & i = n+1\\
		K & & i = n+2
		\end{array}
	\right. 
\\
r^\tinyB_{a(n+2)+i} \;=\; s^\tinyB_{a(n+2)+i} &=&
        a(n+2) +
        \left\{\begin{array}{lcl}
                i+2 &\quad& i = 1,\dots,n \\
                n+4 & & i = n+1\\
		2n+4-K & & i = n+2
                \end{array}
        \right. 
\end{eqnarray*}
and for $a = mJ$ the azure sums are zero and the beige sums are
\begin{eqnarray*}
r^\tinyB_{a(n+2)+i} \;=\; s^\tinyB_{a(n+2)+i} &=&
	a(n+2) +
        \left\{\begin{array}{lcl}
                i+2 &\quad& i = 1,\dots,K \\
                i+1 &     & i = K+1,\dots,n\\
		n+2 &     & i = n+1,n+2.
                \end{array}
        \right.
\end{eqnarray*}
Finally, we define the cyan sums. For $j=0,\dots,J-1$ and $k =
0,\dots,m-1$ let $a = jm+k$ and $b = mJ-1-a$.  If $e_k = (u,v)$
(with $u<v$), then
\begin{eqnarray*}
\begin{array}{lclclcl}
r^\tinyC_{a(n+2)+u} &=& 2
	&\quad&
	s^\tinyC_{b(n+2)+n-u+1} &=& 1 \\
r^\tinyC_{a(n+2)+v} &=& 1
	&\quad&
	s^\tinyC_{b(n+2)+n-v+1} &=& 2 \\
r^\tinyC_{a(n+2)+n+1} &=& 1
	&\quad&
	s^\tinyC_{b(n+2)+n+1} &=& 1. 
\end{array}
\end{eqnarray*}
The row and column sums not defined above are assumed to be zero.

%%%%%%%%%%%%%%%%%%%%%%%%%%%%%%%%%%%%%%%%%%%%%%%%%%%%%%%%%%%%%%

\subsection{Realizations of Azure and Beige Atoms}
%           ------------------------------------

Let $\cal A$ and $\cal B$ be $(n+2) \times (n+2)$ matrices completely
filled with azure and beige atoms, respectively.  We will use notation
$\cal A(\VEC\gamma,\VEC\delta)$ for realizations of
$ASM(\VEC\gamma,\VEC\delta)$ and $\calB(\VEC\alpha,\VEC\beta)$ for
realizations of $BSM(\VEC\alpha,\VEC\beta)$.  We define
$\VEC\pi$ by
\[
	\VEC\pi =
		\underbrace{0\ldots0}_K\underbrace{1\ldots1}_{n-K}.
\]
For 0-1 vectors $\VEC\beta^0, \VEC\alpha^0, \ldots, \VEC\beta^{mJ-1},
\VEC\alpha^{mJ-1}$, each of total sum $n-K$, consider $L \times L$
azure-and-beige matrices of the following form:

%%%%%%%%%%%%%%%%%%%%%%%%%%%%%%%%%%%%%%%%%%%%%%%%%%
%
%
{\small
\newcommand{\A}[2]{\calA(\VEC\alpha^{#1},\VEC\beta^{#2})}
\newcommand{\B}[2]{\calB(\VEC\alpha^{#1},\VEC\beta^{#2})}
\newcommand{\Bmin}{\calB(\VEC\pi,\VEC\beta^{0})}
\newcommand{\Bmax}{\calB(\VEC\alpha^{mJ-1},\VEC\pi)}
\begin{eqnarray}
	\left[
	\begin{array}{ccc@{}cccc}
\calA	& \calA	&\calA 	&\cdots & \A{mJ-1}{mJ-1}& \Bmax        \\[.6em]
\calA	& \calA	&\calA 	& 	& \B{mJ-2}{mJ-1}&\calB		\\[.6em]
\calA	& \calA	&\calA	& 	& \calB		&\calB		\\[-.4em]
\vdots	& 	& 	&	& 		& \vdots	\\
\calA	& \calA	&\A22	& 	& \calB		& \calB		\\[.6em]
\calA 	& \A11	&\B12	& 	& \calB		& \calB		\\[.6em]
\A00 	& \B01	& \calB	& 	& \calB		& \calB		\\[.6em]
\Bmin   &\calB	& \calB	&\cdots & \calB		& \calB		\\[.6em]
	\end{array}
	\right].
	\label{eqn: a/b realization}
\end{eqnarray}
} 
%
%
%%%%%%%%%%%%%%%%%%%%%%%%%%%%%%%%%%%%%%%%%%%%%%%%%%

\begin{lemma}	\label{lem: a/b realization}
Let $\calI^{\tinyAB}$ be the restriction of $\calI$ to the azure and
beige sums only.  Then a matrix $T$ is a realization of
$\calI^{\tinyAB}$ if and only if $T$ has the
form~(\ref{eqn: a/b realization}), where
\begin{equation}				\label{eqn: order}
		\VEC\pi
	\preceq \REV\beta^0
	\preceq	\VEC\alpha^0 
	\preceq \REV\beta^1 
	\preceq	\VEC\alpha^1 
	\preceq \ldots
	\preceq	\REV\beta^{mJ-1}
	\preceq	\VEC\alpha^{mJ-1}
	\preceq \REV\pi.
\end{equation}
\end{lemma}

\begin{Proof}
Note that by Lemma~\ref{lem: beige skew mirror}
and~\ref{lem: azure skew mirror}, a matrix of the
form~(\ref{eqn: a/b realization})
exists iff inequalities (\ref{eqn: order}) are true.

\noindent
$(\Leftarrow)$\/
Let $T$ be a matrix of the form~(\ref{eqn: a/b realization}). 
By straightforward verification of the row and column sums we
obtain that $T$ is a realization of
$\calI^\tinyAB$. (Note that the entries of $\VEC\alpha^i$,$\VEC\beta^i$,
$i = 0,\dots,mJ-1$ appear in the beige sums with the plus sign
in the azure skew mirrors, and with the minus sign in the beige skew mirrors.)

\noindent
$(\Rightarrow)$\/
Let now $T$ be a realization of $\calI^\tinyAB$, and
denote by $F$ an arbitrary matrix of the form~(\ref{eqn: a/b realization}).
Block $(a,b)$ consist of entries in rows $a(n+2)+i$ and columns
$b(n+2)+j$ for all $i,j=1,\ldots,n+2$.  We call it
\[\mbox{
	\begin{tabular}{ll}
	an \emph{upper-left block}		&if $a+b<mJ$,\\
	the \emph{side-diagonal block $b$}	&if $a+b=mJ$,\\
	the \emph{main-diagonal block $b$}	&if $a+b =mJ+1$,\\
	and a \emph{lower-right block}		&if $a+b>mJ+1$.
	\end{tabular}}
\]
We claim that $T$ has the structure depicted in Figure~\ref{fig: decomp}.
\begin{figure}[ht]
\centerline{\begin{tabular}{cc}
 \epsfig{file=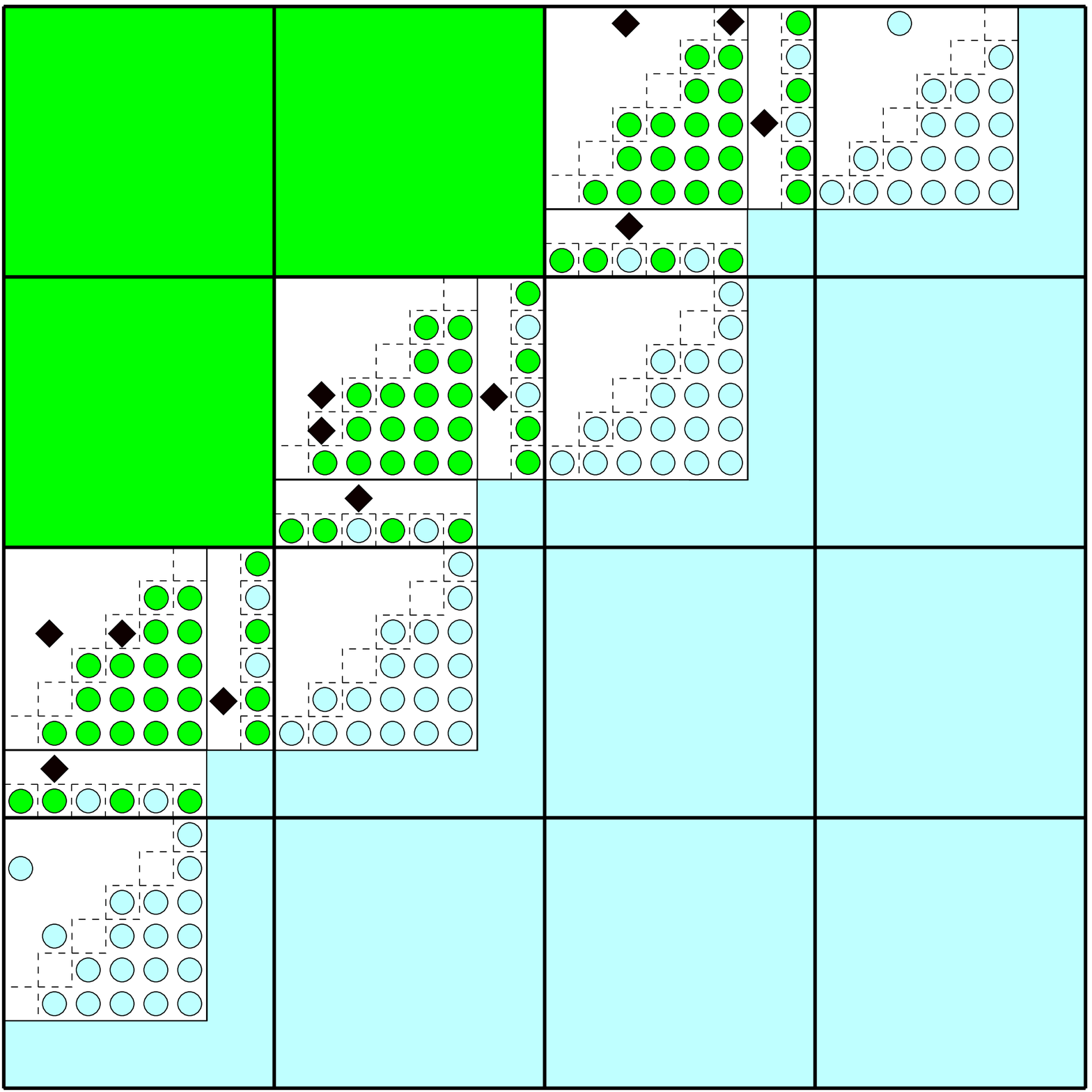,width=8cm}
&\epsfig{file=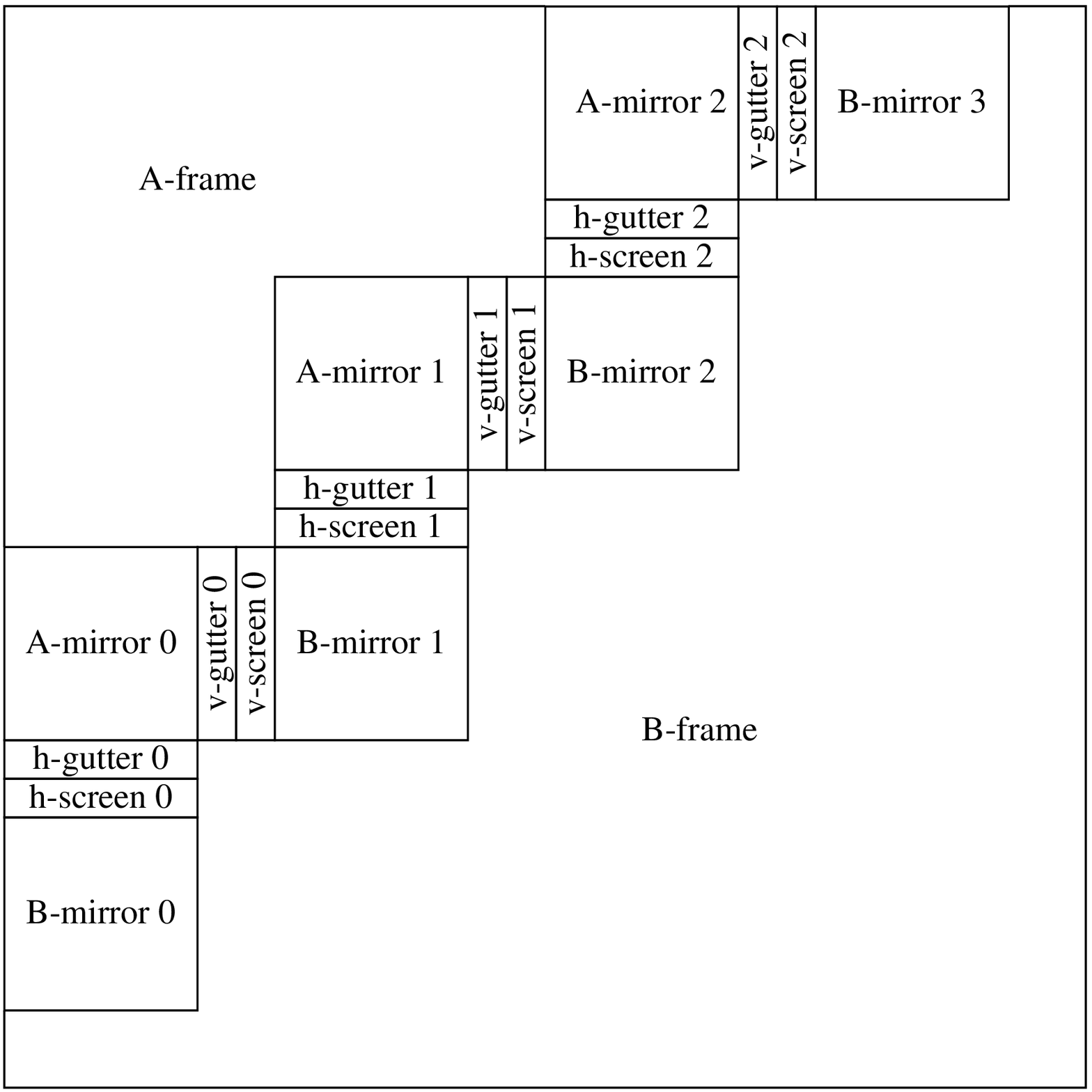,width=8cm}
	    \end{tabular}
}
\caption{\em\small A realization of $\calI$ and its
	abstract structure. The pictures do not show an actual
	example, but rather are meant only to
	illustrate the general structure of a realization (even for our
	small 6-vertex graph, a full matrix would
	have $28\times 28$ blocks). In this particular realization all
	A-mirrors are perfect mirrors associated with
	$\sigma=101011$. This 0-1 vector encodes the vertex set
	$\{1,3,5,6\}$.  The three shown edge verifiers check that
	edges $(3,5)$, $(4,5)$ and $(1,4)$ are covered
	by that vertex set.
}
\label{fig: decomp}
\end{figure}
Let $\barT^\tinyA$ be the 0-1 matrix representing the \emph{non-azure}
cells in $T$: $\barT^\tinyA[i,j] =1$ if and only if $T[i,j]\neq A$.
Let $\barF^\tinyA$ be the analogous matrix for $F$.  Then
$\barT^\tinyA$ and $\barF^\tinyA$ are realizations of the same
instance of 1-CCP.  Therefore, if $F^\tinyA$ is $(k,l)$-decomposed so
must be $T^\tinyA$.  Let $k=(mJ-d)(n+2)$ and $l=d(n+2)$ for any
$d=0,\ldots,mJ$.  Then $\barF^\tinyA$ is $(k,l)$-decomposed.  Thus all
upper-left blocks in $T$ are exactly $\calA$ and all main-diagonal and
lower-right blocks have no azure atoms. The union of all upper-left blocks
will be called the \emph{A-frame}.

Let $T^\tinyB$ be the 0-1 matrix $T$ representing the beige atoms in
$T$, that is $T^\tinyB[i,j]=1$ if and only if $T[i,j] = B$.  Let
$F^\tinyB$ be the analogous matrix for $F$. Let $k,l$ be a pair
of indices such that either $k=(mJ+1-d)(n+2)-2$
and $l=d(n+2)-2$ for some $d=1,\ldots,mJ$, or
$(k,l) \in \braced{(0,L-2),(L-2,0)}$.  Then, by the structure of
realizations of azure and beige skew mirrors in
Lemmas~\ref{lem: beige skew mirror} and \ref{lem: azure skew mirror},
$F^\tinyB$ is $(k,l)$-decomposed,
and so must be $T^\tinyB$.  The region in $T$
corresponding to the $1$'s in the submatrices $\bottomright T_{kl}$
(for the $k,l$ chosen above) is called the \emph{B-frame}.  So the
B-frame of $T$ is all beige. Since the lower-right blocks are included
in this region, they are exactly $\calB$.

We partition the side-diagonal block $b$ into
\[
\mbox{\begin{tabular}{ll}
A-mirror $b$	& upper left $n\times n$ corner\\
B-corner $b$	& lower right $2\times2$ corner\\
v-screen $b$	& remaining entries in column $n+2$\\
h-screen $b$	& remaining entries in row $n+2$\\
v-gutter $b$	& remaining entries in column $n+1$\\
h-gutter $b$	& remaining entries in row $n+1$.
      \end{tabular}
}
\]
	Our previous observation implies that the B-corner is
completely beige (as it is part of the B-frame), and the A-mirror does
not contain any beige atoms. The gutters are completely empty and the
screens completely filled, since
\begin{eqnarray*}
	r^\tinyA_{b(n+2)+n+1} + r^\tinyB_{b(n+2)+n+1} =& L-n&=
	s^\tinyA_{b(n+2)+n+1} + s^\tinyB_{b(n+2)+n+1}\\
	r^\tinyA_{b(n+2)+n+2} + r^\tinyB_{b(n+2)+n+2} =& L&=
	s^\tinyA_{b(n+2)+n+2} + s^\tinyB_{b(n+2)+n+2}.
\end{eqnarray*}

	We define 0-1 vectors $\VEC\beta^0, \VEC\alpha^0, \ldots,
\VEC\beta^{mJ-1}, \VEC\alpha^{mJ-1}$ to represent positions of the
beige atoms in the screens: $\alpha^b_i=1$ if and only if $i$-th atom
(from top) in v-screen $b$ is beige and $\beta^b_i=1$ if and only if
$i$-th atom (from left) in h-screen $b$ is beige.

Since the main-diagonal blocks have no azure atoms, each side-diagonal
blocks $b$ is a realization of $\ASM(\VEC\alpha^{b},
\VEC\beta^{b})$. Moreover, the main-diagonal blocks $0$ and $mJ$ are
realizations of, respectively, $\BSM(\VEC\pi, \VEC\beta^0)$ and
$\BSM(\VEC\alpha^{mJ-1}, \VEC\pi)$, and each other main-diagonal block
$b$, for $b = 1,\dots,mJ-1$, is a realizations of
$\BSM(\VEC\alpha^{b-1},\VEC\beta^{b})$.  Lemmas~\ref{lem: beige skew
mirror} and~\ref{lem: azure skew mirror} imply inequalities~(\ref{eqn:
order}).
\end{Proof}

%%%%%%%%%%%%%%%%%%%%%%%%%%%%%%%%%%%%%%%%%%%%%%%%%%%%%%%%%%%%%%

\subsection{The Correctness Proof}
%           ---------------------

\begin{theorem}
The problem 3-CCP is NP-complete in the strong sense.
\end{theorem}

\begin{Proof}
Clearly, 3-CCP is in NP. To justify the correctness of the reduction
described in Section~\ref{sec: The reduction}, we need to prove that
$G$ has a vertex cover of size $K$ if and only if $\calI$ is consistent.

\begin{description}
\item[$(\Rightarrow)$] Suppose that $U$ is a vertex cover of size $K$
in $G$.  Define $\VEC\gamma$ by $\gamma_u = 0$ iff $u\in U$.  Let $T$
be a matrix of the form (\ref{eqn: a/b realization}) in which
$\VEC\alpha^i = \VEC\gamma$, and $\VEC\beta^i = \REV\gamma$ for $i =
0,\dots,mJ-1$.  We have $\VEC\pi \preceq \VEC\gamma \preceq \REV\pi$.
By Lemma~\ref{lem: a/b realization}, $T$ is a realization of
$\calI^\tinyAB$.  Since $U$ is a vertex cover, Lemma~\ref{lem: edge
verifier} implies that $T$ can be extended to a realization of
$\calI$.
\item[$(\Leftarrow)$]
Let $T$ be any realization of $\calI$.  By
Lemma~\ref{lem: a/b realization}, $T$ restricted to azure and beige
atoms has the form (\ref{eqn: a/b realization}). 
Lemma~\ref{lem: majorization is shallow} implies that the sequence
\begin{eqnarray*}
		\REV\beta^{m0} 
	\preceq \REV\beta^{m1} \preceq 
		\ldots
	\preceq \REV\beta^{m(J-1)}
	\preceq \REV\beta^{mJ},
\end{eqnarray*}
(where $\VEC\beta^{mJ} = \VEC\pi$) has at most $K(n-K)+1$ distinct
vectors, and thus $\REV\beta^{ma} = \REV\beta^{m(a+1)}$ for some $0\le
a \le J-1$. By (\ref{eqn: order}), we get
\begin{eqnarray*}
  \REV\beta^{ma} 	= \VEC\alpha^{ma} = 
  \REV\beta^{ma+1} 	= \VEC\alpha^{ma+1} = 
	  \ldots =
  \REV\beta^{ma+m-1} 	= \VEC\alpha^{ma+m-1}.
\end{eqnarray*}
Define $U = \braced{ u : \alpha^{am}_u = 0}$.  Using
Lemma~\ref{lem: edge verifier}, we obtain that $U$ is a vertex cover.
\end{description}

To complete the proof, note that $\calI$ consists of $6L=O(n^5)$
numbers each bounded by $L$, so the unary encoding of $\calI$
has size $O(n^{10})$. Moreover, this encoding
can be computed in polynomial time. We conclude that
3-CCP is strongly NP-complete. 
\end{Proof}

%%%%%%%%%%%%%%%%%%%%%%%%%%%%%%%%%%%%%%%%%%%%%%%%%%%%%%%%%%%%%%
%%%%%%%%%%%%%%%%%%%%%%%%%%%%%%%%%%%%%%%%%%%%%%%%%%%%%%%%%%%%%%
%%%%%%%%%%%%%%%%%%%%%%%%%%%%%%%%%%%%%%%%%%%%%%%%%%%%%%%%%%%%%%

\section{Final Comments} \label{sec: Final Comments}
%       ----------------

We proved that $c$-CCP is NP-complete for $c\ge 3$.  Since it is known
that $1$-CCP can be solved efficiently in polynomial time (see
\cite{Brualdi80}), the only unresolved case is for $c=2$.

\paragraph*{Relation to multicommodity flows.}
Consider the following problem: given a bipartite
directed graph $H = (U,V,E)$,
where $E$ is the set of arcs directed from $U$ to $V$, with each arc
having capacity 1, we want to ship two commodities from the
vertices in $U$ to the vertices in $V$, according to the given
supplies in $U$ and demands in $V$.  More specifically, for each
vertex $u_i\in U$ we are given a supply $x^a_i$ of commodity $a$, and
for each vertex $v_j\in V$ we are given a demand $y^a_j$ of commodity
$a$, where $a\in\braced{1,2}$.  We wish to compute an integral
2-commodity flow from $U$ to $V$ of maximum total value.  Let us call
it {\em 2-Commodity Integral 2-Layer Flow}, or {\em 2-CI2LF}.  It is
known (see, \cite{EvItSh76}) that the 2-commodity integral flow
problem is NP-hard for directed networks.  We can improve it to the
2-layer case.  By modifying the argument outlined in Section~\ref{sec:
General Idea}, it is not difficult to show that 2-CI2LF is NP-hard as
well: simply note that all but two atom types have unique
realizations which are independent of the given instance $G,K$ of
Vertex Cover, and associate the entries not occupied by these atoms
with the edges of the resulting graph $H$. (Another proof can be
obtained by modifying the proof in \cite{GaGrPr97b} in a similar
fashion.)

The argument above does not imply that 2-CCP is NP-complete, since the
graphs corresponding to the 2-CCP problem are {\em complete\/}
bipartite graphs. This leads to the following open problem: Can
2-CI2LF be solved in polynomial time for complete bipartite graphs?

\paragraph*{Consequences to data security problems.}
Similar to \cite{GaGrPr97b}, our result has some consequences for
problems arising in statistics and data security.

The reconstruction problem for contingency tables is similar to the
1-CCP problem, except that now we allow a realization to contain any
non-negative integers (not just 0's and 1's).  Our result implies that
this problem is NP-hard even when we want to reconstruct a table whose
entries are in the set $\braced{0,1,\mu,\mu^2}$, for some given $\mu$.
(To see this, modify the proof by representing each table entry in a
$\mu$-ary notation, where $\mu = L+1$, and associate color sums with
the coefficients of $1$, $\mu$ and $\mu^2$.)

A related problem, arising in the 3D statistical data security
problem, is to reconstruct a 3D table from its projections, which are
called the row, column and file sums. Irving and Jerrum
\cite{IrvJer94} proved that this problem is NP-hard even when all file
sums are either 0 or 1. The work in \cite{GaGrPr97b} implies that the
problem is NP-hard for $L\times L\times 7$ tables and all file sums
equal 1. Our result improves this result further to
tables of size $L\times L\times 4$.

%\cite{Anstee83,Anstee86,Anstee87,Brualdi80,BruRos80,Chang71,Chen86}
%\cite{CheSas89,Chen92,FiLaReSh91,GaGrPr96,GaGrPr97a,GaGrPr97b}
%\cite{GerSlu82,IrvJer94,KSBSKO95,Nam97,Ryser63,SKSBKO93,EvItSh76}

\bibliographystyle{plain}

\bibliography{discrete_tomography}

\end{document}